\documentclass[aps,pra,twocolumn,showpacs,preprintnumbers,amsmath,amssymb,superscriptaddress,10pt]{revtex4-1}

\usepackage{amsmath,amssymb}
\usepackage{bm}
\usepackage[latin1]{inputenc}
\usepackage{dcolumn}
\usepackage{graphicx}
\usepackage{SIunits}
\usepackage{hyphenat}

\begin{document}

\title{Tailored bright illumination attack on distributed-phase-reference protocols}

\author{Lars Lydersen}
\email{lars.lydersen@iet.ntnu.no}
\author{Johannes Skaar}
\affiliation{Department of Electronics and Telecommunications, Norwegian University of Science and Technology, NO-7491 Trondheim, Norway}
\affiliation{University Graduate Center, NO-2027 Kjeller, Norway}

\author{Vadim Makarov}
\affiliation{Department of Electronics and Telecommunications, Norwegian University of Science and Technology, NO-7491 Trondheim, Norway}

\date{\today}

\begin{abstract}
Detector control attacks on quantum key distribution systems exploit the linear mode of avalanche photodiode in single photon detectors. So far, the protocols under consideration have been the
BB84 protocol and its derivatives. Here we present how bright tailored illumination exploiting the linear mode of detectors can be used to eavesdrop on distributed-phase-reference protocols, such as differential-phase-shift and coherent-one-way.
\end{abstract}


\maketitle

\section{Introduction}
Quantum mechanics theoretically allows two parties, Alice and Bob, to grow a private, secret key, even if the eavesdropper Eve can do anything permitted by the laws of nature \cite{bennett1984,ekert1991,lo1999,shor2000}. The field of quantum key distribution (QKD) has evolved rapidly in the last two decades, with transmission distances reaching $250\,\kilo\meter$ in the laboratory \cite{stucki2009}, and commercial QKD systems available from several vendors \cite{comqkdsystems}.

Even though QKD can be proved secure in theory \cite{mayers1996,lo1999,shor2000}, the implementations may contain loopholes allowing side-channel attacks. In fact, many such side-channel attacks have been identified and countered, either by modifying the implementation, generalizing the security proof, or both \cite{vakhitov2001,gisin2006,makarov2006,qi2007,lamas-linares2007,fung2007,makarov2008,zhao2008,nauerth2009,lydersen2010,xu2010,gottesman2004,koashi2006,fung2009,maroy2010}. The discoveries of implementation loopholes does not prove the insecurity of QKD, but rather its maturity. Scrutinizing the implementations is a vital step to achieve satisfactory security in practical QKD.

Recently the detector control attacks \cite{makarov2009,lydersen2010a,lydersen2010b,wiechers2010,gerhardt2010,sauge} exploiting the linear mode of avalanche photodiodes (APDs) received considerable attention. This class of attacks stands out from previous attacks since they allow the eavesdropper Eve to copy the full key, while not being revealed by monitored parameters, such as the quantum bit error rate. Furthermore, the attacks are implementable with current technology. The loophole has been identified in two commercial QKD systems \cite{lydersen2010a}, and the full attack has been demonstrated under realistic conditions on an experimental QKD setup \cite{gerhardt2010}. Furthermore, it has been proved possible to keep the APDs in the linear mode through blinding illumination for both passively-quenched \cite{makarov2009,gerhardt2010}, actively-quenched \cite{sauge} and gated APDs \cite{lydersen2010a,lydersen2010b,wiechers2010} through a variety of techniques.

So far these bright illumination detector control attacks have been considered on the Bennett-Brassard 1984 (BB84) protocol \cite{bennett1984} and its derivatives with similar implementations, such as the Scarani-Acin-Ribordy-Gisin 2004 (SARG04) \cite{scarani2004}, Ekert \cite{ekert1991}, six-state \cite{bruss1998,bennett1984a}, and decoy protocols \cite{hwang2003,wang2005a,lo2005}. Here we present how to exploit the linear mode of the detectors in distributed-phase-reference protocols such as differential-phase-shift (DPS) \cite{inoue2002,inoue2003} and coherent-one-way (COW) \cite{gisin2004,stucki2005} to tracelessly eavesdrop the full raw and secret key.

\section{Eavesdropping on linear detectors}
For the distributed-phase-reference protocols considered here, the implementation of Bob is `passive' in the sense that Bob does not use a modulator that introduces randomness into his detection system (similarly to passive vs$.$ active basis choice in the BB84 protocol \cite{rarity1994}). Detector control is easier for passive implementations, since Eve does not have to deal with different possibilities associated with Bob randomly selecting a measurement.

\begin{figure}[b!]
  \includegraphics[width=8.6cm]{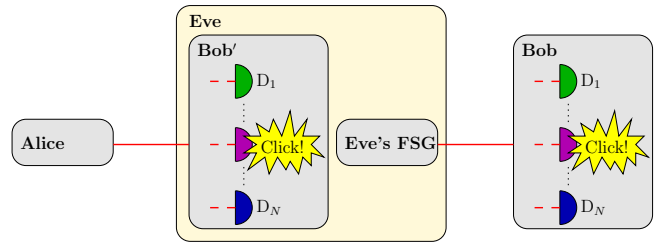}
  \caption{Scheme of a detector control attack: Eve measures the states from Alice using a copy of Bob's measurement device (Bob$'$) to obtain a detection event. She then uses a faked-state generator (FSG) \cite{makarov2005} to generate a bright pulse tailored to cause the same detection event in Bob.}
  \label{fig:detector-control-attack} 
\end{figure}

In a detector control attack, Eve measures the states from Alice using a copy of Bob's measurement device (Bob$'$) to obtain a detection event. Eve resends a bright trigger pulse targeting the detectors operated in the linear mode, and in a successful attack it causes the exact same detection event in Bob's measurement device (see Fig.~\ref{fig:detector-control-attack}). Since Eve uses a copy of Bob, Bob's detection statistics will be indistinguishable from the detection statistics he obtains without any eavesdropper. Therefore, Alice's and Bob's data will not reveal the eavesdropper. Furthermore, since Eve has an exact copy of Bob's detection results, the details of the protocol are irrelevant: the security is broken for any classical post-processing since Eve can listen to the classical channel and perform the same post-processing on her copy of the detection results. Therefore, the challenge in a detector control attack is to find a way to make arbitrary detection events in Bob's measurement device, given that the detectors are accessible in the linear mode.

If the detectors are gated, it may be possible to access them in the linear mode simply by sending the bright states after the gate \cite{wiechers2010}. Otherwise it might be necessary to blind the detectors, either with continuous illumination \cite{lydersen2010a,lydersen2010b} or different types of modulated blinding \cite{lydersen2010b,sauge}.

Regardless of how the linear mode is obtained, the detectors have similar characteristics and have two important parameters: $P_{\text{never}}$ is the maximum trigger pulse power which never causes a click in any detector, and $P_{\text{always}}$ is the minimum trigger pulse power which always causes a click in an arbitrary detector. For BB84 \cite{bennett1984}, SARG04 \cite{scarani2004} and decoy-protocols \cite{hwang2003,wang2005a,lo2005} the requirement for perfect detector control attacks is given by \cite{lydersen2010a}
\begin{equation}
  P_{\text{always}} < 2P_{\text{never}}.
  \label{eq:dps-req}
\end{equation}
Note that both $P_{\text{never}}$ and $P_{\text{always}}$ seem to increase with higher blinding illumination \cite{lydersen2010a}.

\section{Differential-Phase-Shift}
\begin{figure}[t!]
  \includegraphics[width=8.6cm]{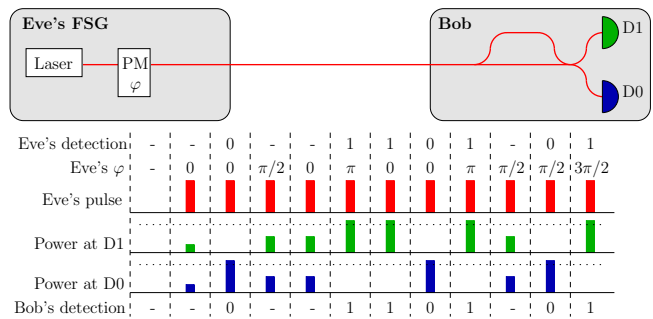}
  \caption{Detector control in a DPS implementation: Eve's FSG consists of a laser source producing a train of coherent pulses with amplitude $P_{\text{always}}$, and a phase modulator (PM). Eve controllably causes identical detection events in Bob by an appropriate phase $\varphi$ in each pulse. No trigger pulse is applied to the left of the diagram, nor in the first slot.}
  \label{fig:dps} 
\end{figure}

The upper right of Fig.~\ref{fig:dps} shows Bob's measurement device in the DPS protocol \cite{inoue2002,inoue2003}, consisting of an unbalanced Mach-Zehnder interferometer and one detector for each bit value. The length difference of the arms in the interferometer matches the time difference between two adjacent pulses sent by Alice. Alice sends a train of coherent pulses and uses the phase difference between two adjacent pulses to encode the two different bit values: 0 phase difference corresponds to the bit value 0 and $\pi$ phase difference corresponds to the bit value 1.

For the DPS protocol, Eve's faked-state generator (FSG) is simply a copy of Alice's optical scheme, but the coherent pulses are brighter with amplitude $P_{\text{always}}$. As we will see, the requirement for the detection thresholds is the same as for the BB84 family of protocols, given by Eq.~\ref{eq:dps-req}. Assuming suitable detection thresholds, an arbitrary detector at Bob in slot $k$ can be triggered by selecting the phase difference $\varphi_k - \varphi_{k-1}$:
\begin{equation*}
  \varphi_k - \varphi_{k-1} = 
  \begin{cases}
    (N+1/2)\pi  &\text{causes a vacuum event,}\\
    2N\pi       &\text{causes a click in D0,}\\
    (2N+1)\pi   &\text{causes a click in D1,}
  \end{cases}
\end{equation*}
where $N$ is an integer (see Fig.~\ref{fig:dps}). Since $P_{\text{always}}/2$ hits each detector for vacuum events, the requirement for perfect eavesdropping is given by Eq.~\eqref{eq:dps-req}.

If Bob accepts at least two vacuum events between every detection event, for instance due to low transmission in the quantum channel or detector deadtime, Eve can relax the requirement \eqref{eq:dps-req} further. Then she does not send a train of pulses, but rather for each detection event she sends two pulses with the appropriate phase difference. Then $P_{\text{always}}/4$ hits the detectors in the slots before and after the detection event.

If blinding is necessary, one can use a blinding light source with a coherence length less than the length difference of the interferometer arms to illuminate both detectors equally.

\vspace{-.09cm}
\section{Coherent-One-Way}
\begin{figure}[t!]
  \includegraphics[width=8.6cm]{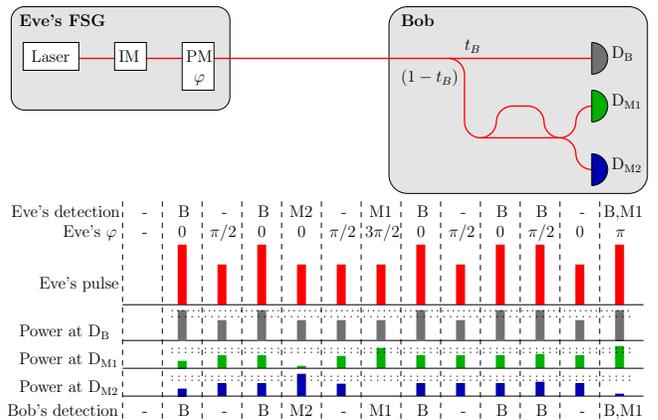}
  \caption{Detector control in a COW implementation: Eve's FSG consists of a laser source producing a train of coherent pulses, an intensity modulator (IM) and a phase modulator (PM). The lower part of the figure shows an example of detector control, where the system parameters are assumed to be $t_B = 0.5$, $P_{\text{never,M}} = 400\,\micro\watt$, $P_{\text{always,M}} = 500\,\micro\watt$, $P_{\text{never,B}} = 600\,\micro\watt$ and $P_{\text{always,B}} = 750\,\micro\watt$. The dashed lines in the diagram are the detector thresholds. Eve controllably causes identical detection events in Bob by an appropriate phase $\varphi$ and amplitude in each pulse. The trigger pulse amplitude is $2P_{\text{always,M}} = 1000\,\micro\watt$, and is increased to $2P_{\text{always,B}} = 1500\,\micro\watt$ to trigger the data detector. No trigger pulse is applied to the left of the diagram, nor in the first slot.}
  \label{fig:cow} 
\end{figure}

The upper right of Fig.~\ref{fig:cow} shows Bob's measurement device in the COW protocol \cite{gisin2004,stucki2005}, consisting of a fiber-optic coupler or a beam-splitter with splitting ratio \nohyphens{$t_B$~:~$(1-t_B)$,} followed by a detector D$_\text{B}$ to generate secret key. The other output of the fiber-optic coupler leads to an unbalanced interferometer identical to the one in the DPS protocol. It has two monitoring detectors D$_\text{M1}$ and D$_\text{M2}$ to check for eavesdropping, by checking the coherence of adjacent pulses. To generate key, Alice sends a train of pulses which are grouped in pairs. In each pair, the slot absent of a pulse determines the bit value of the key. The exact details of the protocol are irrelevant, as long as Eve can transparently mirror her detection events onto Bob's detectors.

As we will see, for perfect eavesdropping against COW, it is necessary for Eve to obtain different trigger pulse thresholds for the data detector D$_\text{B}$, and the monitoring detectors D$_{\text{M0}}$ and D$_{\text{M1}}$. $P_{\text{always,B}}$ and $P_{\text{never,B}}$ are the thresholds for the data detector while $P_{\text{always,M}}$ and $P_{\text{never,M}}$ are the thresholds for the monitoring detectors.

The monitoring detectors have exactly the same setup as for DPS, and therefore they can be controlled as described in the previous section. However, since only $(1-t_B)$ of the trigger pulse power enters the interferometer of the monitoring detectors, the amplitude of the pulse train must be increased to $P_{\text{always,M}}/(1-t_B)$. For perfect control, it is important that the illumination which enters the other arm does not trigger the data detector. This requires that
\begin{equation}
  \frac{t_B}{1-t_B} P_{\text{always,M}} < P_{\text{never,B}}.
  \label{eq:cow-req-1}
\end{equation}
The data detector can be triggered by increasing the amplitude of the trigger pulse to $P_{\text{always,B}}/t_B$. Then however, it is crucial that the monitoring detectors are not triggered. To minimize the illumination on the monitoring detectors, phase difference is set to $\pi/2$, and the threshold requirement is given by
\begin{equation}
  \frac{1-t_B}{t_B} P_{\text{always,B}} < 2P_{\text{never,M}},
  \label{eq:cow-req-2} 
\end{equation}
where the factor 2 represents that the illumination is split between the two monitoring detectors, just as the factor appearing in Eq.~\eqref{eq:dps-req}. Again, if there are at least two vacuum events between every detection event, the factor 2 can be replaced by 4. Furthermore, some COW implementations use only one monitoring detector \cite{stucki2008,stucki2009}. In that case the requirement~\eqref{eq:cow-req-2} can be relaxed even further, since most of the illumination can be directed to the unused interferometer output during vacuum events.

\begin{table}[b!]
  \caption{Data detector thresholds for various values of $t_B$, given by Eq.~\ref{eq:cow-reqs} for $P_{\text{never,M}} = 400\,\micro\watt$ and $P_{\text{always,M}} = 500\,\micro\watt$.} 
  \begin{tabular*}{5cm}{@{\extracolsep{\fill}} c c c}
      $t_B$ & $P_{\text{never,B}} >$ & $P_{\text{always,B}} <$  \\ \hline
      0.5  &  500$\,\micro\watt$  &   800$\,\micro\watt$ \\
      0.8  & 2000$\,\micro\watt$  &  3200$\,\micro\watt$ \\
      0.9  & 4500$\,\micro\watt$  &  7200$\,\micro\watt$ \\
      0.95 & 9500$\,\micro\watt$  & 15200$\,\micro\watt$ \\
  \end{tabular*}
  \label{tab:thresholds}
\end{table}

To see what the requirements \eqref{eq:cow-req-1} and \eqref{eq:cow-req-2} means in practice, they can be rewritten to
\begin{subequations}
  \begin{equation}
     P_{\text{never,B}} > \frac{t_B}{1-t_B} P_{\text{always,M}}, \\
  \end{equation}
  \begin{equation}
     P_{\text{always,B}} < 2\frac{t_B}{1-t_B} P_{\text{never,M}}.
  \end{equation}
  \label{eq:cow-reqs}
\end{subequations}
Now let us assume the reasonable values \cite{lydersen2010a,lydersen2010b} $P_{\text{never,M}} = 400\,\micro\watt$, and that $P_{\text{always,M}} = 500\,\micro\watt$ for the monitoring detectors. Table~\ref{tab:thresholds} lists different constraints on the data detector thresholds for various values of $t_B$, where $t_B = 0.5$ and $t_B = 0.9$ have been reported in experiments \cite{stucki2005,stucki2008,stucki2009}. With $t_B$ close to 1, the thresholds for the data detector must be very much higher than for the monitoring detectors. If blinding illumination is applied, the data detector will receive a larger fraction of the illumination which would usually cause higher thresholds \cite{lydersen2010a} than for the monitoring detectors. If a fiber-optic coupler is used, Eve may increase the threshold difference even further by using a blinding wavelength outside the working range of the coupler, blinding the data detector even deeper \footnote{This can easily be countered with a wavelength filter at Bob's entrance. However any extra optical component added to Bob increases photon loss and thus reduces the key generation rate and maximum communication distance.}. In the implementations with only one monitoring detector, Eve can control the amount of blinding illumination at it independently from the data detector, by splitting the blinding illumination arbitrarily between the output ports of the interferometer. However, how much higher thresholds it is possible to achieve for the data detector than the monitoring detectors remains an open question. Note that for values of $t_B$ close to 1, it should be straightforward for Eve to trigger the data detector while keeping the monitoring detectors silent. Therefore, it is important to check for the absence of clicks in the monitoring detectors.

Eve's FSG is nearly a copy of Alice's optical scheme; in addition it includes a phase modulator to control the monitoring detectors. Eve emits a train of coherent pulses with amplitude $P_{\text{always,M}}/(1-t_B)$. To make the data detector click, the amplitude is increased to $P_{\text{always,B}}/t_B$. One of the monitoring detectors in slot $k$ can be triggered by selecting the phase difference $\varphi_k - \varphi_{k-1}$:
\begin{equation*}
  \varphi_k - \varphi_{k-1} = 
  \begin{cases}
    (N+1/2)\pi  &\text{causes a vacuum event,}\\
    2N\pi       &\text{causes a click in D$_{\text{M2}}$,}\\
    (2N+1)\pi   &\text{causes a click in D$_{\text{M1}}$,}
  \end{cases}
\end{equation*}
where $N$ is an integer. Since the data detector and the monitoring detectors are controlled independently, it is straightforward to cause a simultaneous click in the data detector and one of the monitoring detectors. Figure~\ref{fig:cow} shows Eve's FSG, and an example of detector control.

\section{Discussion and Conclusion}
We have derived the requirements \eqref{eq:dps-req} and \eqref{eq:cow-reqs} for a perfect attack on DPS and COW assuming that Eve must introduce a click deterministically in Bob. However, if the line between Alice and Bob is lossy, Eve might place her Bob$'$ closer to Alice, and receive more detections than Bob would expect. To simulate this loss, instead of applying $P_{always}$, Eve can reduce the power in the trigger pulses to a level which triggers the detector with a probability equal to the expected transmittance. This relaxes the requirements \eqref{eq:dps-req} and \eqref{eq:cow-reqs}.

Since the detector threshold requirement \eqref{eq:dps-req} has been obtained using numerous techniques \cite{lydersen2010a,lydersen2010b,wiechers2010}, the DPS protocol is obviously vulnerable to the bright illumination attack. For the COW protocol, the requirements \eqref{eq:cow-reqs} on the detector thresholds depend on the splitting ratio between the data and monitoring detectors. Routing more light to the data detector increases the required difference in detection thresholds. It remains an open question for which splitting ratios suitable detector thresholds can be obtained. However, it seems that the bright illumination attacks represent a significant threat to the security of the COW protocol, and all subsequent implementations should be investigated thoroughly.

While several countermeasures have been proposed \cite{makarov2009,lydersen2010a,lydersen2010b,wiechers2010,gerhardt2010,yuan2010}, none of them have been proved secure \cite{lydersen2010c} to our knowledge. The same countermeasures should apply to the distributed-phase-reference protocols. A frequently mentioned countermeasure is a power meter at Bob's entrance. As long as this countermeasure has not been proven secure, it has to be considered insufficient \cite{lydersen2010b,lydersen2010c}. Nevertheless, it makes life harder for Eve.

We have shown that distributed-phase-reference protocols DPS and COW are vulnerable to bright tailored illumination attacks. This emphasizes the generality of these attacks, and demonstrates the importance of scrutinizing all implementations and protocols thoroughly, as this is a vital step for obtaining suitable practical security for QKD.

\begin{acknowledgments}
We acknowledge useful discussions with Nicolas Gisin, Hugo Zbinden, Nino Walenta, Christoffer Wittmann, and Valerio Scarani. This work was supported by the Research Council of Norway (grant no$.$ 180439/V30).
\end{acknowledgments}

\end{document}